# A Hybrid Model to Extend Vehicular Intercommunication V2V through D2D Architecture

Emad Abd-Elrahman[*], Adel Mounir Said[*], Thouraya Toukabri[**], Hossam Afifi[***] and Michel Marot[***]

[*]National Telecommunication Institute, Cairo, Egypt; [**]Orange Labs, Issy Les Moulineaux, France; [***]Institut Mines-Telecom, RST Department, Saclay, France.

{emad.abd_elrahman, hossam.afifi, michel.marot}@telecom-sudparis.eu, amounir@nti.sci.eg, thouraya.toukabrigunes@orange.com

*Abstract*— In the recent years, many solutions for Vehicle to Vehicle (V2V) communication were proposed to overcome failure problems (also known as *'dead-ends'*). This paper proposes a novel framework for V2V failure recovery using Device-to-Device (D2D) communications. Based on the unified Intelligent Transportation Systems (ITS) architecture, LTE-based D2D mechanisms can improve V2V *'dead-ends'* failure recovery delays. This new paradigm of hybrid V2V-D2D communications overcomes the limitations of traditional V2V routing techniques. According to NS2 simulation results, the proposed hybrid model decreases the end to end delay (E2E) of messages delivery. A complete comparison of different D2D use cases (best & worst scenarios) is presented to show the enhancements brought by our solution compared to traditional V2V techniques.

## I. INTRODUCTION

Pure Vehicular communications also known as V2V can suffer from blocking or failure issues called *'dead-ends'* [1]. This is mainly due to geographical guidance errors, insufficient communication range, or topology dynamicity. Hence, unrelayed V2V messages will block the whole communication process. The V2V communication is used either to disseminate local information in the VANET domain or to relay it to the nearest Road Side Unit (RSU), which will transmit it to ground servers and Traffic control Centers (TCC). The Intelligent Transportation Systems (ITS) communication architecture standardized by ETSI and ISO [2] has been proposed in this context as a global and harmonized architecture that enhances V2V routing mechanisms. The ITS reference architecture provides V2V failover mechanisms through vertical handover mechanisms provided by the embedded ITS protocol stack of vehicles' communication systems. Meanwhile, new architectures and mechanisms enabling D2D communications in cellular networks have become a hot discussion and an uprising research topic as they enable a new type of short range direct communications based on LTE. This new type of communication based on devices proximity information is envisioned to promote for a variety of new generation services going from the real-time location-based and context-aware services to public safety services. Recently standardized at the 3GPP level [3], LTE-assisted D2D communications [4] are considered as an open door to cooperative communications with heterogeneous networks including ITS. Yet, the idea of combining LTE architecture capabilities with V2X communications has been addressed in a recent study [5] in the "LTE4V2X" framework. This study proposes a self-organizing architecture and a centralized vehicular network organization over the LTE network. These timely advances already give a glance at the next generation of smart vehicles, which would be equipped with a variety of access technologies (LTE, IEEE802.11p, etc.) for reliable and more efficient vehicular communications.

The work presented in this paper is inspired from the ITS architecture principles and the new LTE-based D2D communication mechanisms to offer a cognitive model solving V2V blocking and failure issues. The idea is to exploit the extended ITS network management features allowing vertical handovers between different access media to recover V2V dead-ends. We believe that a mixed architecture combining V2V and intermediate D2D communication improves the overall transmission success ratio and delay. The D2D support can be seen as a failover recovery solution that could be a little bit slower than direct V2V in the worst case (discovery phase is done on demand), or which help interconnecting disconnected groups of mobile nodes and enhance the processing delay in the best case (discovery phase is done proactively).

The paper is organized as follows; Section II discusses related works and the recent advances in ITS and D2D, Section III introduces our hybrid mechanism for V2V failure recovery based on D2D and a generic routing algorithm. Simulations results are explained in Section IV for different schemes under different constraints (best & worst cases) and are compared to a traditional V2V mechanism based on GPSR protocol. Finally, the conclusion and the future work directions are discussed in Section V.

## II. RELATED WORKS

### A. ITS Reference Architecture

The reference ITS communication architecture [2] is the fruit of about ten years of harmonization and standardization efforts on the definition of a uniform communication model for next generation transportation systems. This architecture would be deployed on various types of ITS stations involved in Cooperative ITS communications (i.e. vehicle ITS station, personal ITS station, central ITS station and roadside ITS station), but different features would be supported according to the type of ITS station, deployment environment and user needs. This reference architecture offers the flexibility to deploy on the same platform a variety of ITS applications with different requirements [6] [7].

As depicted in Fig. 1, the ITS architecture is a protocol stack inspired from the OSI model. It is composed of vertical and horizontal layers, each of them composed of protocol and software sets. Software communications between horizontal and vertical layers are provided through Service Access Points (SAPs) which are logical interfaces that allow the exchange of cross layer commands and information. Two new vertical planes (management and security) are further introduced to provide cross layer security and management services to all ITS layers for better performance and efficiency. For instance, the management plan is designed to store the ITS station configuration parameters and collect information from all the horizontal layers in order to provide a decision support system that adapt the network routing decisions with the service level requirements and optimize the distribution of services on the ITS network. This enables vertical handovers between different access media that could overcome transmission failure cases and maintain sessions continuity using D2D Communications.

*B. D2D communications*

D2D communications as an underlay to cellular networks have been the focus of many recent research' works [8] [9] [10]. Defined as a direct communication between devices in range proximity without the involvement of a network infrastructure, D2D opens the opportunity for a new generation of proximity-based services including services for offloading, connectivity extension, network capabilities enhancement, public safety as well as real-time location-based and context-aware services. Recently, network-assisted D2D schemes has gained more attention due to their advantages in providing trust, privacy and security functions as well as Quality of Service (QoS) mechanisms through mobile operators networks. As a result, a technical specification on ProSe (Proximity Services) has been recently defined by the 3GPP (3$^{rd}$ generation partnership project) in order to define the necessary extensions to the LTE architecture for the support of D2D communications, including pure direct D2D discovery and communication mechanisms as well as network-assisted ones.

*C. Recent Contributions in the Dead-end Issue*

D2D and ITS communications are somehow similar as they are based on the discovery of proximate neighbors through a short range access technology, thus, enabling mobile ad-hoc communications. For instance, an ITS personal station [7] is a D2D-enabled device using a specific D2D service dedicated to ITS. Then, the ITS personal station capable of D2D communication (i.e. using LTE direct media) would discover its D2D neighbors through the LTE network and then play the role of a heterogeneous relay to ensure the transmission of V2V information to a wider distance in poor V2V coverage situations. As a matter of fact, existing studies on both D2D and V2V communications are lacking from such mechanisms.

Our vision in this paper is to introduce this cooperative hybrid model combining V2V and LTE-assisted D2D communications' capabilities to provide an efficient and performant solution for failure recovery in V2V. The D2D mechanism used in our hybrid model is described in Section III and is inspired from [10].

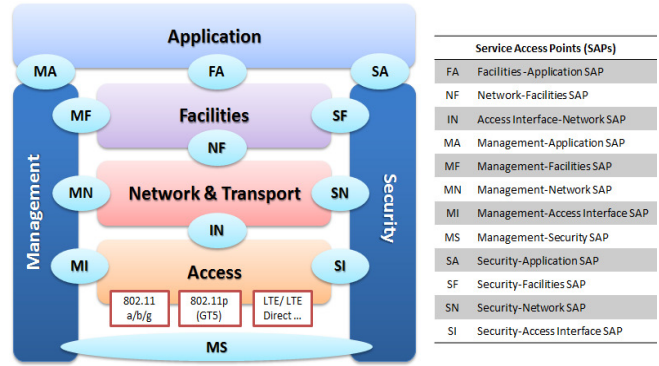

Fig. 1. ITS station reference architecture [2].

In VANET, some contributions study messages relaying and spreading copies over networks [11]. Others studies address the routing issue from a geographical aspect using geo-location and GPS [12]. The problem of *'dead ends'* in VANET happens when the routing disconnects due to a *'hole to next hop'*. In this case a compensating mechanism can be used to help V2V packet routing along the alternative path. Proposed ways depend on either go-back *'one'* or *'many hops'* to find another relay for messages. Other solutions tried to overcome this problem by redirecting the packet in the reverse way in order to find an inverse path to the RSU. Many ad-hoc protocols details deal with the *'dead ends'* problems and are explained in [13]. Also, parallel paths are considered as backup solutions for V2V routing in case of failure [1] [14].

In addition to solving the *'hole to next hop'* problem, the proposed contribution in this paper details a new genericV2V routing model for failure recovery where Broadcast / Directed / Relaying, and Back / Relaying / Inverse techniques are used. Crosschecked models with simulations confirm the robustness of our hybrid protocol and the strength of using generic models instead (or in addition) of simulations with GPSR protocol. The order in which these two ideas are presented is inversed because we want to follow a logical scenario highlighting the methodology before jumping to the solution.

III. HYBRID COMMUNICATIONS

Typically, V2V applications for safety and infotainment are proposed based on IEEE802.11p [15] wireless communications. Several solutions are available in the literature based on routing or broadcast. For simplicity, we chose the broadcast model although routing or Q-learning could be used alternatively. The message relaying will stop when the destination is reached or when a RSU takes over.

Otherwise, the message delivery fails when any of the two conditions is not reached ('*dead-ends*').

### A. Generic V2V Routing Recovery, Problem Statement

In this section, our V2V Routing Approach (V2V-RA) is introduced. V2V-RA is used to recover the unrelayed messages due to the '*dead-ends*' in VANETs as shown in Fig. 2 (the red point).

We suppose that in the case of failure, the vehicle (*x*) at a given position has no neighbors to relay the alert message that it is received from its predecessor. V2V-RA is proposed to resolve this problem as follows:

1) The vehicle (x) will return the message in the backward direction to inform the previous vehicle (x-1) about its position as a *dead-end*, and hence the message cannot be forwarded in that direction (i.e. vehicle (x) has no neighbor vehicles to forward the alert message to it).

2) Vehicle (x-1) has three sequencing options to solve the *dead-end* situations before considering that the message has failed:

    a) First, vehicle (x-1) will check in its neighbor table for other vehicles.
        o If vehicle *(x-1)* has another neighbor, it will redirect the alert message to that node.
        o Else, it continues in a recursive manner.
    b) Second, it will look for another RSU that could be reached on the road. This would mandates to change the direction of the transmission relaying.
    c) If the previous two options are not available, the recovery algorithm steps are repeated starting from step (1) considering vehicle (x-1) instead of (x). i.e. going back by one hop (which is considered as worst case).

In the second option, the transmission direction is changed to another RSU location using a geographical routing protocol. One possible method is to use our previous work [16] [17] based on beamforming-based relaying, or Q-learning-based routing as proposed in [18].

In this work, we concentrate our comparison on using geographical routing protocols to find an alternative path in case of failure. In the simulation part, our proposed model results are compared with the geographical routing protocol GPSR (Greedy Perimeter Stateless Routing for Wireless Networks) [19] to validate the efficiency of the D2D solution.

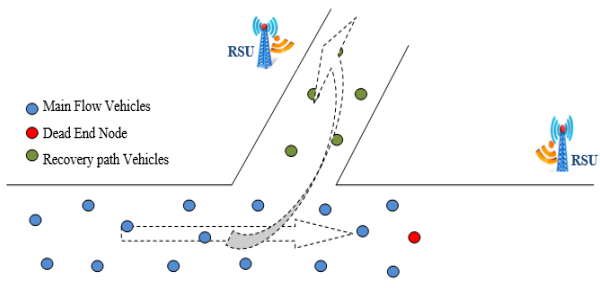

Fig. 2. V2V-RA Model.

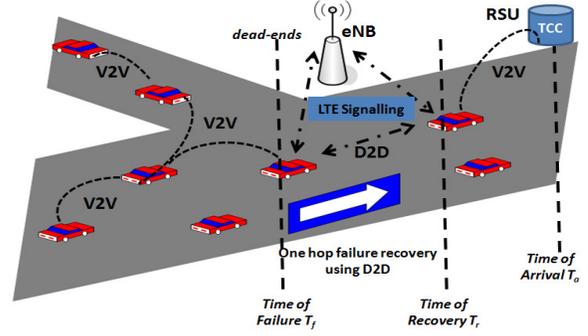

Fig. 3. The hybrid model of V2V assisted by D2D.

### B. Hybrid Solution

Our solution combines ITS-adapted V2V concept with the D2D LTE-assisted architecture for its globalization and reliability. This could be considered as a concrete example of cognitive radio use case. As shown in Fig. 3, when a '*dead-end*' occurs ($T_f$), the ITS protocol stack of the last V2V transmitting node can make a channel sensing followed by vertical handover to a D2D transmission mechanism. The D2D protocol exchanges are explained in detail in [10] and we briefly sketch it hereafter for the sake of comprehension. D2D implies the presence of two devices in the range proximity and attached to the same LTE base station eNB (evolved Node B) or adjacent ones; the D2D-enabled devices can discover each other using different methods as explained in [8], [9], and [10]. The strength of this solution is that the V2V coverage area extension. Thanks to the LTE-Direct transmission used in the D2D mechanism [10] in which a D2D-enabled device can discover devices up to 1 km in its surrounding. This increases the chances to reach a new V2V hop belonging to a next connected set of vehicles in the direction of the final destination. The Overhead Time (OHT) based on the D2D mechanism could obey to the following equation:

$$D2D_{OHT} = T_{\text{D2D Discovery Time}} + 2*T_{\text{handover time from .11p to D2D}} + T_{V2V(L \text{ rest of road till RSU})}$$

The values of these times will be considered according to the D2D use cases explained in the results section. Hence, from this short analysis, we can conclude that the D2D discovery time is determinant to achieve a performant hybrid V2V-D2D failure recovery solution.

### C. D2D Neighbor Discovery

The D2D discovery phase described hereafter is presented with more details in [10]. One important step in our hybrid model is the discovery of the adequate next D2D-enabled vehicle to avoid the V2V transmission failure. In the LTE Radio Access Network (RAN), a vehicle is seen as a D2D-enabled device. We assume that it is under an eNB coverage and that it is in the same direction of information relaying. We also assume that the D2D-enabled device embedded on the vehicle's communication system is using an ITS specific D2D service. The communication between the D2D stack and the V2V service is made through the ITS station protocol stack introduced in Section II/A. Hence, it represents a use case implementation of cognitive radio principles.

In the D2D process, devices in range proximity need first to discover each other. According to the model proposed in [10], D2D-enabled devices are firstly authenticated and authorized by the LTE core network to use D2D radio resources. This pre-discovery step is assisted by the LTE network through the eNB: it provides mainly a reliable basis for the discovery and communication between devices. Then, a direct discovery phase is done between D2D-enabled devices through an LTE-direct interface using presence beacons.

*D. D2D Scenarios*

Our solution's performance is very sensitive to the discovery phase. We propose two methods for its evaluation:

- *Proactive V2V&D2D*: In this scenario, we perform the standard discovery phase for direct D2D communication before the V2V failure. This means that the discovery time is Zero ($T_D=0$) and the handover from V2V to D2D is done immediately after the failure occurs ($T_{HO} = 0$).

- *On-demand V2V&D2D*: In this scenario, we propose that the standard discovery phase for D2D communication is done on demand (i.e. triggered by the failure detection). This means that the discovery time ($T_D$) is considered as an overhead in the overall processing and the handover time ($T_{HO}$) as well.

## IV. SIMULATIONS & COMPARISON

In this section, we will evaluate our hybrid V2V-D2D solution through simulation against V2V-RA. We conducted all simulation studies using NS2. The simulations include two scenarios as follows:

*A. V2V Assisted by D2D Communications*

D2D-enabled terminals are continuously monitoring their vicinity zone to discover proximate devices that are available to establish a D2D link for communication (as shown in Fig. 3). The communication range can vary from 500, 1000 up to 1500m depending on communication zone obstacles. Also, the D2D direct discovery is done in LTE by scanning predefined channels, which gives us a very short latency time compared to wireless IEEE 802.11*xx* discovery methods.

In this part, we suppose that the V2V failure occurs 4 kilometers far from the nearest RSU. Our objective is to transfer an alert message to this RSU. The simulation considers six sub-cases as follows for different road lengths (L) defined by the proposed D2D communication ranges:

- Proactive V2V&D2D (Best-Case 1.5km): discovery time ($T_D$ = Zero) and the handover from V2V to D2D, $T_{HO}$ = Zero. The road length left ($L_{Rest}$) from the total road length (4 kilometers) is equal to 2.5km based on a D2D-Direct communication over 1.5km (i.e. $L_{Rest}$ = 2.5km).

- Proactive V2V&D2D (Best-Case 1km): The same assumption as previous one except that $L_{Rest}$ becomes 3km.

- Proactive V2V&D2D (Best-Case 500m): The same assumptions as previous one except that $L_{Rest}$ = 3.5km.

- On-demand V2V&D2D (Worst-Case 1.5km): Discovery time ($T_D$) and handover ($T_{HO}$) exist. The road length left from the total road length is equal to 1km (i.e. $L_{Rest}$ = 2.5km for D2D communication range 0.5km).

- On-demand V2V&D2D (Worst-Case 1km): With $L_{Rest}$ = 3km for D2D communication range 1km.

- On-demand V2V&D2D (Worst Case 0.5km): With $L_{Rest}$ = 3.5km for D2D communication range 500m.

*B. V2V Assisted by Traditional Broadcast Routing*

In this scenario, we consider a traditional solution for node recovery in normal V2V ad-hoc communications. It is compared to the previous D2D mechanism based on the same road length (4Km) until the RSU and following two scenarios:

- V2V-RA (Best-Case): In this case, we succeed to find an alternative route after one hop backward.

- V2V-RA (Worst-Case): In this case, we assume that the routing algorithm needs more than one hop to find a route to the RSU.

In our previous work [20], we did complete analytical and simulation evaluations for generic V2V routing. The connected set model gave reasonable results for different communication ranges and with different vehicle densities. Also, this model is compared to existing V2V routing models in generic way. But, this paper will focus on the comparison with a standard V2V routing protocol called GPSR.

The NS2 simulation parameters [21] are listed in Table I. We assumed in this work that the LTE-D2D service is always enabled in vehicles or could be triggered on-demand. Moreover, the D2D links are available as an output for Public Safety (PS) Proximity Service (ProSe) needs.

Table I: Simulation parameters.

| Parameters | Settings |
| --- | --- |
| Communication Ranges | 250, 350, 550m |
| Vehicle Density | From 20 to 100 |
| Road Length | From 1 to 4 km |
| Vehicle Speed | From 30 to 100 km/h |
| Packet Size | 256-512 KB |
| Packet Rate | 0.5 s |
| Simulation Time | 600 s |
| Mobility Model | IDM : VanetMobiSim [22] |

Fig.4 shows a comparison between the GPSR protocol and the LTE-assisted D2D in terms of the number of hops. The GPSR communication ranges are 250m and 350m, and the D2D communication range is 550m. It is clear that, the number of hops in the D2D case is less than in the GPSR cases (V2V communication). Another advantage for D2D is the stability in number of hops regardless the vehicle densities.

Fig. 5 shows the same comparison but in terms of the consumed delay along the path starting from the failure recovery time ($T_f$) until reaching the nearest RSU ($T_a$). We noticed that, although the number of hops may be constant for different vehicle densities, the E2E delay increased due to the effects of radio interferences (i.e. more radio interfaces due to dense vehicles will lead to more delay consumption for message transfer). The lowest curve for D2D indicates the less consumed time for transferring alarms in PS applications using D2D ProSe.

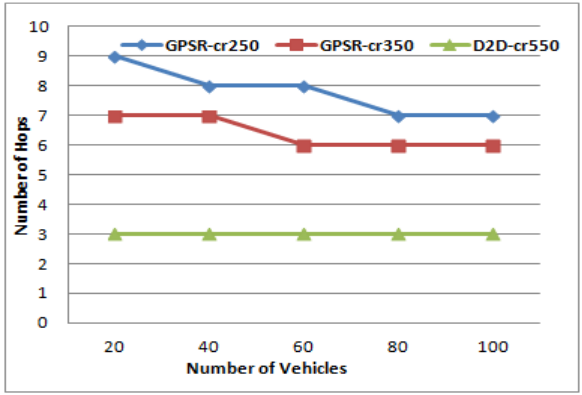

Fig. 4. NS2 simulation for number of hops along with 2Km path with different number of vehicles in case of different communication ranges for GPSR simulation againest D2D.

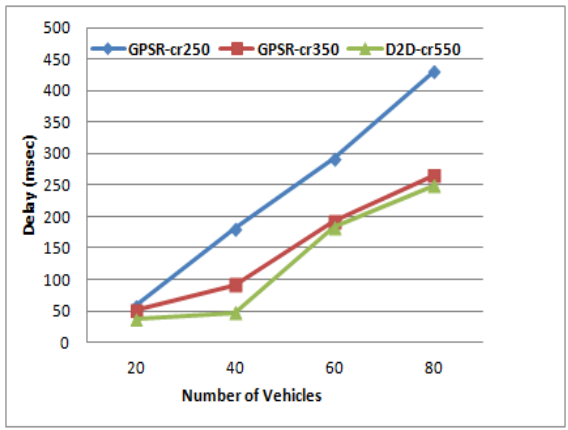

Fig. 5. NS2 simulation for delay consumed on 2Km path starting from recovery point till final RSU.

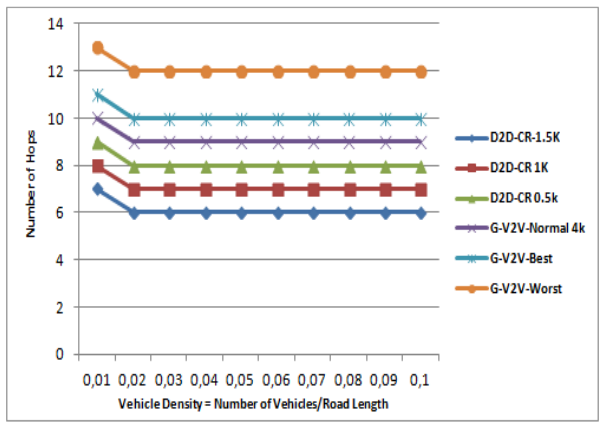

Fig. 6. Simulation results for D2D use cases with different communication range (CR) versus pure V2V routing recovery in terms of number of hops.

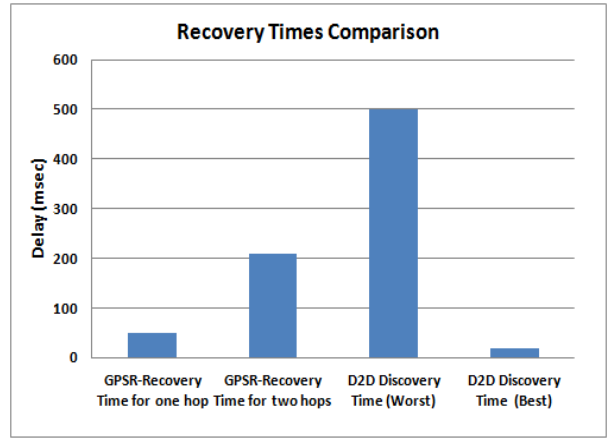

Fig. 7. Recovery times for GPSR simulation against D2D discovery time.

The 3GPP [3] proposed different performance metric values for D2D discovery and communication evaluations. They considered that the D2D discovery does not exceed 64 msec (according to Qualcomm simulations [23]). Hence, the latency time for the discovery phase could be negligible. Moreover, they considered a 100 msec per hop for relays as a system level metric for communication. So, if we consider D2D direct as a one hop relay for our alert message in a Public Safety (PS) service, we have an overhead time of about 100 msec to be considered as the overall delay in the communication chain. Compared to a one hop recovery in pure V2V routing of 50 ms average time as estimated in [24], it seems that at a first glance that V2V gives a shorter delay. As shown in Fig. 6, this is not true as the overall E2E delay in the D2D case will be reduced compared to V2V, regardless to the number of backward hops. This is explained by the fact that the D2D communication range (LTE based) is always 3 to 5 times larger than the V2V range (IEEE 802.11 family). This confirms that our solution improves delay over pure V2V for one or more hops failure recovery.

*C. General Comparisons*

According to our queuing model for the D2D discovery time [25], we used the best case for time consumed in discovery (10 msec) and the worst case of 500 msec as shown in Fig. 7. Also, we compare these times to the recovery time of the GPSR protocol. The recovery time scenarios suppose the best case is almost (50 msec) and for the worst case is (200 msec). It is clear from this figure that D2D gives the best use case over all the other cases. Thus, D2D helps achieving a faster recovery in V2V.

The V2X solutions could be merged with the D2D mechanisms for a better cost efficiency either in the ITS essential cost (CAPEX) or operations cost (OPEX). Table II summarizes a cost comparison between the two methodologies and our hybrid solution. It is clear that our hybrid solution has long-term economical benefits and efficiency. In addition to the cost advantages for the hybrid solution, the use of D2D with V2V can have the following advantages in Public Safety (PS) services:

- Better security based on the standard LTE security native features
- Better support for QoS depending on network capabilities
- Connectivity areas extension due to long communication range of LTE compared to IEEE802.11p
- Faster messages transmissions

Table II: Cost comparison.

|  | **Pure V2X** | **Pure D2D** | **Hybrid (V2X-D2D** |
|---|---|---|---|
| **CAPEX** | highly proportional and increased in terms of network growth | depend on LTE infrastructure and 4G network cost | benefit from existing V2X investments besides new LTE 4G deployments |
| **OPEX** | separate network for public safety needs more operations cost | use the mobile network and is included in cellular operational cost | hybrid operation can achieve balance between public safety network and 4G networks |

## V. CONCLUSIONS

In this paper, we presented a novel framework for V2V failure recovery using LTE-Direct D2D assisted solution. This hybrid solution enhances the public safety proximity service performance by minimizing the overall E2E delay along the communication path. Moreover, different scenarios of D2D discovery either proactive or on-demand were compared to the standard GPSR protocol scenario using NS2 simulation. Simulation results proved the efficiency and the performance of the vehicular ad-hoc network when a D2D mechanism is used to recover V2V communication failures. In the future, we will consider more sophisticated scenarios of D2D in case of a disaster happened to eNB. We expect a delay overhead for discovery plus communication without eNB.

As our proposed solution requires each vehicle to operate D2D and GPSR simultaneously, the future work will focus more on this operation details especially; interference, energy consumption and handover for this cognitive communication.